\documentclass{article}
\usepackage[utf8]{inputenc}
\usepackage[english]{babel}
 
\usepackage{amsmath}
\usepackage{amssymb}
\usepackage{amsmath}
\usepackage{amsthm}
\usepackage{dsfont}
\usepackage{footnote}
\usepackage{hyperref}
\usepackage[algo2e]{algorithm2e}
\usepackage{algorithm}
\usepackage{algorithmic}
\usepackage{pifont}
\usepackage{comment}

\newtheorem{theorem}{Theorem}

\newcommand{\cmark}{\ding{51}}%
\newcommand{\xmark}{\ding{55}}%

\title{SEA-BREW: A Scalable Attribute-Based Encryption Scheme for Low-Bitrate IoT Wireless Networks}
\author{Michele La Manna $^{\star}$ $^{\dagger}$, Pericle Perazzo $^{\dagger}$, Gianluca Dini $^{\dagger}$.}
\date{$^{\star}$ University of Florence, $^{\dagger}$ University of Pisa (DII)}

\usepackage{natbib}
\usepackage{graphicx}

\begin{document}

\maketitle
\begin{abstract}
Attribute-Based Encryption (ABE) is an emerging cryptographic technique that allows one to embed a fine-grained access control mechanism into encrypted data.
In this paper we propose a novel ABE scheme called SEA-BREW (Scalable and Efficient Abe with Broadcast REvocation for Wireless networks), which is suited for Internet of Things (IoT) and Industrial IoT (IIoT) applications. 
In contrast to state-of-the-art ABE schemes, ours is capable of securely performing key revocations with a single short broadcast message, instead of a number of unicast messages that is linear with the number of nodes.
This is desirable for low-bitrate Wireless Sensor and Actuator Networks (WSANs) which often are the heart of (I)IoT systems.
In SEA-BREW, sensors, actuators, and users can exchange encrypted data via a cloud server, or directly via wireless if they belong to the same WSAN.
We formally prove that our scheme is secure also in case of an untrusted cloud server that colludes with a set of users, under the generic bilinear group model.
We show by simulations that our scheme requires a constant computational overhead on the cloud server with respect to the complexity of the access control policies.
This is in contrast to state-of-the-art solutions, which require instead a linear computational overhead.

\end{abstract}

\sloppy
\section{Introduction}
\label{sec:introduction}

In the Internet of Things (IoT) vision \cite{atzori2010internet,gilchrist2016industry,sicari2015security,granjal2015security}, ordinary ``things'' like home appliances, vehicles, industrial robots, etc. will communicate and coordinate themselves through the Internet.
By connecting to Internet, things can provide and receive data from users or other remote things, both directly or via cloud.
Cloud-based services are in turn provided by third-party companies, such as Amazon AWS or Microsoft Azure, usually through pay-per-use subscription.
On the other hand, outsourcing sensitive or valuable information to external servers exposes the data owner to the risk of data leakage.
Think for example of an industrial IoT network that communicates and processes business-critical information.
A data leakage could expose a company or an organization to industrial espionage, or it can endanger the privacy of employees or customers.
Encrypting data on cloud servers is a viable solution to this problem.
An emerging approach is \textit{Attribute-Based Encryption} (ABE) \cite{sahai2005fuzzy,goyal2006attribute,bethencourt2007ciphertext,yu2010achieving,rasori2018ABE,yu2011fdac}, which is a cryptographic technique that embeds an access control mechanism within the encrypted data.
ABE describes data and decrypting parties by means of \emph{attributes}, and it regulates the ``decryptability'' of data with \emph{access policies}, which are Boolean formulas defined over these attributes.
In ABE, encrypting parties use an \textit{encryption key}, which is public and unique, whereas any decrypting party uses a \textit{decryption key}, which is private and different for each of them.

Unfortunately, state-of-the-art ABE schemes are poorly suitable for the majority of IoT applications.
The biggest problem is not computational power as one may think, since ABE technology and elliptic curve operations have proven to be well-supportable by mobile devices \cite{ambrosin2015feasibility,ambrosin2016feasibility} and modern IoT devices \cite{girgenti2019feasibility,sowjanya2020elliptic}.
The most problematic aspect is the recovery procedure in case of key compromise, which requires to send an update message to all the devices \cite{yu2010achieving}.
Sending many update messages could be quite burdensome for wireless networks with a limited bitrate, like those employed in IoT \cite{farrell2018low,montenegro2007rfc}.
Indeed modern IoT networks use low-power communication protocols like Bluetooth LE, IEEE 802.15.4, and LoRA, which provide for low bitrates (230Kbps for BLE~\cite{tosi2017performance}, 163Kbps for 802.15.4~\cite{latre2005maximum}, 50Kbps for LoRA~\cite{georgiou2017low}).

In this paper, we propose \textit{SEA-BREW} (Scalable and Efficient ABE with Broadcast REvocation for Wireless networks), an ABE revocable scheme suitable for low-bitrate Wireless Sensor and Actuator Networks (WSANs) in IoT applications. 
SEA-BREW is highly scalable in the number and size of messages necessary to manage decryption keys.
In a WSAN composed of $n$ decrypting nodes, a traditional approach based on unicast would require $\mathcal{O}(n)$ messages. 
SEA-BREW instead, is able to revoke or renew multiple decryption keys by sending a single broadcast message over a WSAN.
Intuitively, such a message allows all the nodes to locally update their keys.
For instance, if $n=50$ and considering a symmetric pairing with 80-bit security, the traditional approach requires $50$ unicast messages of 2688 bytes each, resulting in about 131KB of total traffic. 
SEA-BREW instead, requires a single 252-byte broadcast message over a WSAN.
Also, our scheme allows for per-data access policies, following the \emph{Ciphertext-Policy Attribute-Based Encryption} (CP-ABE) paradigm, which is generally considered flexible and easy to use \cite{bethencourt2007ciphertext,liu2013white,ambrosin2015feasibility}.
In SEA-BREW, things and users can exchange encrypted data via the cloud, as well as directly if they belong to the same WSAN.
This makes the scheme suitable for both remote cloud-based communications and local delay-bounded ones.
The scheme also provides a mechanism of \emph{proxy re-encryption} \cite{yu2010achieving,yu2010attribute,zu2014new} by which old data can be re-encrypted by the cloud to make a revoked key unusable.
This is important to retroactively protect old ciphertexts from revoked keys.
We formally prove that our scheme is adaptively IND-CPA secure also in case of an untrusted cloud server that colludes with a set of users, under the generic bilinear group model.
Furthermore, it can also be made adaptively IND-CCA secure by means of the Fujisaki-Okamoto transformation \cite{fujisaki1999secure}.
We finally show by simulations that the computational overhead is constant on the cloud server, with respect to the complexity of the access control policies.

The rest of the paper is structured as follows. 
In Section \ref{sec:rw} we review the current state of the art.
In Section \ref{sec:system_model} we explain our system model; furthermore, we provide a threat model, the scheme definition, and the security definition for SEA-BREW. 
In Section \ref{sec:procedures} we show the SEA-BREW system procedures.
In Section \ref{math} we mathematically describe the SEA-BREW primitives, and we also show the correctness of our scheme.
In Section \ref{sec:proof} we formally prove the security of SEA-BREW.
In Section \ref{sec:simulations} we evaluate our scheme both analytically and through simulations.
Finally, in Section \ref{sec:conclusion} we conclude the paper.

	\section{Related Work}
	\label{sec:rw}

	In 2007 Bethencourt et al. \cite{bethencourt2007ciphertext} proposed the first CP-ABE scheme, upon which we built SEA-BREW.
	Since then, attribute-Based Encryption has been applied to provide confidentiality and assure fine-grained access control in many different application scenarios like cloud computing \cite{ming2011efficient,yu2010achieving,xu2012dynamic,hur2013improving}, e-health \cite{picazo2014secure}, wireless sensor networks \cite{yu2011fdac}, Internet of Things \cite{touati2015batch,singh2015secure}, smart cities \cite{rasori2018ABE}, smart industries \cite{lamanna2019fabelous}, online social networks \cite{jahid2011easier}, and so on.
	
	With the increasing interest in ABE, researchers have focused on improving also a crucial aspect of any encryption scheme: key revocation.
	In the following, we show many ABE schemes that features different key revocation mechanisms, so that we can compare SEA-BREW to them. 
	First, we recall the notions of \textit{direct} and \textit{indirect} revocation, introduced by \cite{attrapadung2009attribute}.
	Direct revocation implies that the list of the revoked keys is somehow embedded inside each ciphertext.
	In this way, only users in possession of a decryption key which is not in such a list are able to decrypt the ciphertext.
	Instead, indirect revocation implies that the list of the revoked keys is known by the key authority only, which will release some updates for the non-revoked keys and/or ciphertexts.
	Such updates are not distributed to the revoked users.
	In this way, only users that have received the update are able to decrypt the ciphertexts.
	
In table \ref{tab:rw} we provide a summarized visual comparison of SEA-BREW with other schemes.
In the comparison we highlight the following features: (i)\textit{``Immediate Key Revocation"} which is the ability of a scheme to deny -at any moment in time- access to some data for a compromised decryption key; (ii) \textit{``Revocation Type"}, which can be either direct or indirect; (iii) \textit{``Re-Encryption"}, which is the ability of a scheme to update an old ciphertext after a revocation occurs; and (iv) \textit{``Broadcast WSAN Update"}, which is the ability of a scheme to revoke or renew one or more keys with a single message transmitted over a WSAN.
	
\begin{table*}[h]
\resizebox{\textwidth}{!}{%
\begin{tabular}{|c|c|c|c|c|c|c|}
\hline
Schemes & Immediate Key Revocation & Revocation Type & Re-Encryption & Broadcast WSAN Update \\ \hline
Liu et al.\cite{liu2018time}& \cmark         & Direct          & \xmark                &\xmark\\ \hline
Attrapadung et al.\cite{attrapadung2009attribute}&\xmark$\backslash$\cmark& Indirect$\backslash$Direct&\xmark$\backslash$\xmark     & \cmark$\backslash$\xmark\\ \hline
Touati et al.\cite{touati2015batch}&\xmark      & Indirect        & \xmark                & \xmark\\ \hline
Bethencourt et al. ``Naive"\cite{bethencourt2007ciphertext}& \cmark         & Indirect          & \xmark                &\xmark\\ \hline
Cui et al.\cite{cui2016server}  &\cmark      & Indirect        & \xmark        & \xmark\\ \hline
Qin et al.\cite{qin2017server}  &\cmark      & Indirect        & \xmark        & \xmark\\ \hline
Yu et al.\cite{yu2010achieving}&\cmark      & Indirect        & \cmark                & \xmark\\ \hline
SEA-BREW& \cmark                   & Indirect        & \cmark                & \cmark\\ \hline
\end{tabular}%
}
\caption{A summary of prominent ABE schemes that provide a revocation mechanism.}
\label{tab:rw}
\end{table*}

	The scheme of Bethencourt et al. \cite{bethencourt2007ciphertext} lacks functionalities for key revocation and ciphertext re-encryption, which we provide in our scheme.
	However, a naive indirect key revocation mechanism can be realized on such a scheme, but it requires to send a new decryption key for each user in the system, resulting in $\mathcal{O}(n)$ point-to-point messages where $n$ is the number of users.
	In contrast, SEA-BREW is able to revoke or renew a decryption key by sending a single $\mathcal{O}(1)$-sized broadcast message over a wireless network, and it also provides a re-encryption mechanism delegated to the untrusted cloud server.
	
	Attrapadung et al. \cite{attrapadung2009attribute} proposed an hybrid ABE scheme that supports both direct and indirect revocation modes, hence the double values in the associated row of table \ref{tab:rw}. 
    According to the authors, this flexibility is a great advantage to have in a system, because the devices can leverage the quality of both approach depending on the situation.
    The indirect revocation mechanism is based on time slots.
    When a key revocation is performed in the middle of a time slot, it is effective only from the beginning of the next time slot, therefore revocation is not immediate.
    Instead, their direct mechanism implies also the immediate key revocation.
    Notably, with their indirect revocation mechanism, it is possible to revoke or renew a decryption key by sending a single broadcast message over a WSAN.
    However, such message is usually $\mathcal{O}(log(n))$-sized where $n$ is the amount of the users in the system, including the ones revoked in the past.
    Moreover their scheme does not provide any mechanism of re-encryption, therefore if a revoked user somehow is able to get an old ciphertext, he/she is still able to decrypt it.
    Instead, SEA-BREW is able to revoke or renew a decryption key by sending a single $\mathcal{O}(1)$-sized broadcast message, and it also provides a re-encryption mechanism. 
	
	Liu et al. \cite{liu2018time} proposed a Time-Based Direct Revocable CP-ABE scheme with Short Revocation List.
    Since the revocation is direct, the revocation list is embedded in the ciphertext, therefore achieving immediate key revocation.
    Furthermore, the authors managed to condense the entire revocation list in few hundreds bytes, as long as the number of total revocation does not overcome a threshold value.
    However, since the revocation list is destined to grow uncontrollably over time, they propose also a secret key time validation technique.
    This technique allows a data producer to remove a compromised decryption key from the revocation list once such a decryption key has expired.
    Unlike SEA-BREW, this scheme does not provide re-encryption of old ciphertexts.
    Furthermore, the direct revocation mechanism implies that each data producer must know the revocation list. 
    In fact, in SEA-BREW, data producers encrypt their data without knowing any information about revoked consumers.
	
	Touati et al. \cite{touati2015batch} proposed an ABE system for IoT which implements an indirect key revocation mechanism based on time slots.
	In their work, time is divided in slots, and policies can be modified only at the beginning of a slot.
	This approach is efficient only if key revocations and policy changes are known a priori.
	An example is an access privilege that expires after one year.
	Unfortunately, in many systems there is not the possibility to know beforehand when and which access privilege should be revoked.
	For example, in case a decryption key gets compromised the system must revoke it as soon as possible.
	Our scheme gives this possibility.

    Cui et al. \cite{cui2016server}, and Qin et al. \cite{qin2017server} proposed two indirect revocable CP-ABE schemes which do not require to communicate with data producers during a revocation process. 
    However, their schemes require all data producers to be time-synchronised in a secure manner.
    This could be quite difficult to achieve and hard to implement in a WSAN where data producers are often very resource constrained sensors.
    Their schemes do not provide a re-encryption mechanism nor an efficient key update distribution, unlike SEA-BREW.
    Furthermore, SEA-BREW has not the constraint of a tight time synchronization.
    
	Yu et al. \cite{yu2010achieving} proposed an ABE scheme to share data on a cloud server.
	The scheme revokes a compromised decryption key by distributing an update to non revoked users.
	The update is done attribute-wise: this means that only users that have some attributes in common with the revoked key need to update their keys.
	Such update mechanism provides indirect and immediate key revocation, as well as ciphertext re-encryption.
	Notably, their revocation mechanism is not efficient for WSAN, as it requires $\mathcal{O}(n)$ different messages where $n$ is the number of decrypting parties that need to be updated.
	On the other hand, SEA-BREW is able to revoke or renew a decryption key by sending a single $\mathcal{O}(1)$-sized broadcast message over the wireless network.
	
	Finally, from the table, we can see that the scheme proposed by Yu et al. \cite{yu2010achieving} is the one with the most features similar to SEA-BREW.
    Indeed, we will compare the performance of SEA-BREW and the scheme in \cite{yu2010achieving} in section \ref{sec:simulations}
	
\section{System Model and Scheme Definition}
\label{sec:system_model}

\begin{figure*}[t]
	\centering
	\includegraphics[width=\textwidth]{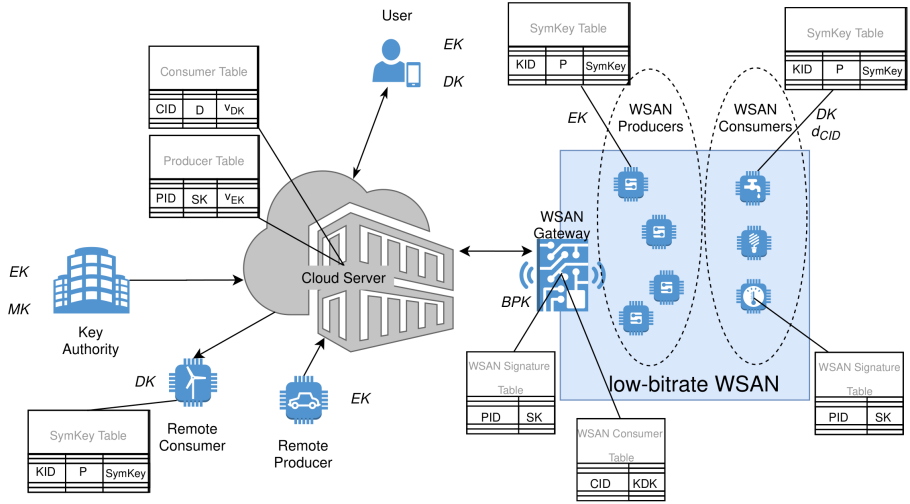}
	\caption{
	SEA-BREW system model. 
	}
	\label{fig:sm}
\end{figure*}

Figure \ref{fig:sm} shows our reference system model.
We assume a low-bitrate WSAN, composed of a set of sensors and actuators, which upload and download encrypted data to/from a \emph{cloud server}.
Sensors and actuators access the cloud server through an Internet-connected \emph{WSAN gateway} node, belonging to the WSAN.
Sensors and actuators inside the WSAN can also communicate directly, without passing through the cloud server.
We assume that some sensors and some actuators are outside the WSAN, and they can also upload and download encrypted data to/from the cloud server, but they cannot communicate directly.
In addition, human users outside the WSAN can upload and download encrypted data to/from the cloud server.
The encrypted data received by an actuator could be a command that the actuator must execute, as well as a measurement from a sensor that the actuator can use to take some decision.
The cloud server is an always-on-line platform managed by an untrusted third-party company which offers storage and computational power to privates or other companies. 
Finally, a fully trusted \emph{key authority} is in charge of generating, updating and distributing cryptographic keys.

In the following, we will call \emph{producers} all those system entities that produce and encrypt data.
This includes sensors internal or external to the WSAN, which sense data, as well as users that produce data or commands for actuators.
Similarly, we will call \emph{consumers} all those system entities that decrypt and consume data.
This includes actuators internal or external to the WSAN, which request data and which receive commands, as well as users that request data.
For the sake of simplicity, we keep the ``producer'' and the ``consumer'' roles separated, however SEA-BREW allows a single device or a single user to act as both.
Producers that are inside the WSAN will be called \emph{WSAN producers}, while those outside the WSAN will be called \emph{remote producers}.
Similarly, consumers that are inside the WSAN will be called \emph{WSAN consumers}, while those outside the WSAN will be called \emph{remote consumers}.

As an use-case example, consider a smart factory with many sensors and actuators which must communicate in a delay-bounded way to implement a real-time application \cite{chen2009real}.
Given the strict requirements, sensors and actuators must communicate directly through the WSAN, without losing time in remote communications with the cloud.
The WSAN inside the smart factory use IEEE 802.15.4 as a link-layer protocol, which is low-energy and low-bitrate.
As a consequence, communications and key management operations must be as lightweight as possible.
In addition, employees, external sensors and external actuators involved for remote applications will upload or download data to/from the cloud server.

Each producer encrypts data by means of an \emph{encryption key} ($\mathit{EK}$).
Each consumer decrypts data by means of a \emph{decryption key} ($\mathit{DK}$).
The encryption key is public and unique for all the producers, whereas the decryption key is private and specific of a single consumer.
A single piece of encrypted data is called \emph{ciphertext} ($\mathit{CP}$).
Each consumer is described by a set of attributes ($\gamma$), which are cryptographically embedded into its decryption key.
The access rights on each ciphertext are described by an \emph{access policy} ($\mathcal{P}$).
We assume that the key authority, the cloud server, and the WSAN gateway have their own pair of asymmetric keys used for digital signature and encryption (e.g., RSA or ECIES keys). 
In addition, each producer and each consumer has a unique identifier called, respectively, \emph{producer identifier} ($\mathit{PID}$) and \emph{consumer identifier} ($\mathit{CID}$), which are assigned by the key authority.
If a device acts as both producer and consumer, then it will have both a producer identifier and a consumer identifier.

When a decryption key needs to be revoked (e.g., because it is compromised or because a consumer has to leave the system), the key authority must ensure that such a decryption key will not be able to decrypt data anymore. 
This is achieved by \emph{Proxy Re-Encryption} (PRE). 
Re-Encryption consists in modifying an existing ciphertext such that a specific decryption key can no longer decrypt it.
This is important to retroactively protect old ciphertexts from revoked keys.
In SEA-BREW, as in other schemes \cite{yu2010achieving}, the Re-Encryption is ``proxied'' because it is delegated to the cloud server, which thus acts as a full-resource proxy for the producers. 
Therefore, data producers do not have to do anything to protect data generated before a revocation.
The cloud server, however, re-encrypts blindly, that is without accessing the plaintext of the messages.
This makes our scheme resilient to possible data leakage on the cloud server.
Our PRE mechanism is also ``lazy'', which means that the ciphertext is modified not immediately after the key revocation, but only when it is downloaded by some consumer. 
This allows us to spread over time the computational costs sustained by the cloud server for the PRE operations. 
We implement the lazy PRE scheme by assigning a version to the encryption key, to each decryption key, and to each ciphertext.
When a key is revoked, the key authority modifies the encryption key, increments its version, and uploads some update quantities to the cloud server.
The set of these update quantities is called \emph{update key}.
The update key is used by the cloud server to blindly re-encrypt the ABE ciphertexts and increment their version before sending them to the requesting consumers.
The cloud server also uses the update key to update the encryption key used by producers, and the decryption keys used by consumers.
Inside the low-bitrate WSAN, instead, the update of the WSAN consumers' decryption keys is achieved with a \emph{constant-ciphertext broadcast encryption scheme}, like the one shown in Boneh et al.'s work \cite{boneh2005collusion}.
The broadcast encryption scheme allows the WSAN gateway to broadcast the update key encrypted in such a way to exclude one or more WSAN consumers from decrypting it.
To do this, the WSAN gateway uses a \emph{broadcast public key} ($\mathit{BPK}$), and each WSAN consumer uses its own \emph{broadcast private key} ($\mathit{d_{CID}}$).
Table \ref{list} lists the symbols used in the paper.

\begin{table}
\centering
\begin{tabular}{|ll|}
\hline
$\mathit{EK}$ & Encryption key \\ \hline
$\mathit{MK}$ & Master key \\ \hline
$\mathit{DK}$ & Decryption key \\ \hline
$\mathit{KDK}$ & Key distribution key \\ \hline
$\mathit{PID}$ & Producer identifier \\ \hline
$\mathit{SK}$ &  Signature verification key \\ \hline
$\mathit{CID}$ & Consumer identifier \\ \hline
$\mathit{KID}$ & Symmetric key identifier \\ \hline
$\mathit{SymKey}$ & Symmetric key \\ \hline
$\mathcal{P}$ & Access policy \\ \hline
$\gamma$ &  Attribute set \\ \hline
$\mathit{BPK}$ & Broadcast public key \\ \hline
$\mathit{d_{CID}}$ & Broadcast private key \\ \hline
$\mathit{CP}$ & Ciphertext \\ \hline
$\mathit{U}$ & Update key \\ \hline
$\mathit{M}$ & Message \\ \hline
\end{tabular}%
\caption{Table of Symbols}
\label{list}
\end{table}

\subsection{Threat Model}
\label{sec:threat}
In this section, we model a set of adversaries and we analyze the security of our system against them.
In particular, we consider the following adversaries: (i) an \emph{external adversary}, which does not own any cryptographic key except the public ones; (ii) a \emph{device compromiser}, which can compromise sensors and actuators to steal secrets from them; (iii) a set of \emph{colluding consumers}, which own some decryption keys; and (iv) a \emph{honest-but-curious cloud server} as defined in \cite{yu2010achieving,rasori2018ABE,di2007over}, which does not tamper with data and correctly executes the procedures, but it is interested in accessing data.
We assume that the honest-but-curious cloud server might collude also with a set of consumers, which own some decryption keys.
Note that the honest-but-curious cloud server models also an adversary capable of \emph{breaching} the cloud server, meaning that he can steal all the data stored in it.
In order to do this, he can leverage some common weakness, for example buffer overflows or code injections, 
or hardware vulnerabilities like Meltdown or Spectre \cite{reidy2018complex}.
We assume that who breaches the cloud server only steals data and does not alter its behavior in correctly executing all the protocols, basically because he tries to remain as stealth as possible during the attack.
Note that this reflects real-life attacks against cloud servers\footnote{\url{https://www.bbc.com/news/technology-41147513}}.
In the following we analyze in detail each adversary model.

The external adversary aims at reading or forging data.
To do so, he can adopt several strategies.
He can impersonate the key authority to communicate a false encryption key to the producer, so that the data encrypted by said producer will be accessible by the adversary. 
This attack is avoided because the encryption keys are signed by the key authority. 
Alternatively, the external adversary can act as a man in the middle between the key authority and a new consumer during the decryption key distribution.
The attacker wants to steal the consumer's decryption key, with which he can later decrypt data. 
This attack is avoided because the decryption key is encrypted by the key authority with asymmetric encryption.
Using the encryption key, which is public, the external adversary may also try to encrypt false data and upload it to the cloud server.
This attack is avoided because he cannot forge a valid signature for the encrypted data, thus he cannot make the false data be accepted as valid by the legitimate consumers. 
To sum up, the external adversary cannot access legitimate data neither inject malicious data.

The device compromiser can compromise a producer or a consumer.
If he compromises a producer, then he gains full control of such a device and full access to its sensed data and to its private key used for signatures.
He cannot retrieve any data sensed before the compromise, because the producer securely deletes data after having uploaded it to the cloud server.
Nonetheless, he can indeed inject malicious data into the system, by signing it and uploading it to the cloud server, or by transmitting it directly to WSAN consumers if the compromised producer belongs to the WSAN.
When the key authority finds out the compromise, it revokes the compromised producer. 
After that, the compromised producer cannot inject malicious data anymore because the private key that it uses for signatures is not considered valid anymore by the consumers.
On the other hand, if the adversary compromises a consumer, then he gains full access to its decryption key.
The attacker can decrypt \emph{some} data downloaded from the cloud server or, if the compromised a consumer belonging to the WSAN, transmitted directly by WSAN producers.
Notably, the adversary can decrypt only data that the compromised consumer was authorized to decrypt.
When the key authority finds out the compromise, it revokes the compromised consumer. 
After that, the compromised consumer cannot decrypt data anymore. 
The reason for this is that our re-encryption mechanism updates the ciphertexts as if they were encrypted with a different encryption key.

A set of colluding consumers can try by combine somehow their decryption keys to decrypt some data that singularly they cannot decrypt.
However, even if the union of the attribute sets of said decryption keys satisfies the access policy of a ciphertext, the colluding consumers cannot decrypt such a ciphertext.
In Section \ref{sec:proof} we will capture this adversary model with the Game 1, and we will provide a formal proof that SEA-BREW is resistant against it.

The honest-but-curious cloud server does not have access to data because it is encrypted, but it can access all the update keys and part of all the consumers' decryption keys.
The update keys alone are useless to decrypt data because the cloud server lacks of a (complete) decryption key.
However, if the cloud server colludes with a set of consumers, then it can access all the data that the consumers are authorized to decrypt.
Interestingly, if the honest-but-curious cloud server is modelling an adversary capable of breaching the cloud server, recovering the breach is easy.
It is sufficient that the key authority generates a new update key, without revoking any consumers.
This has the effect of making all the stolen update keys useless.
On the other hand, in case of an \emph{actual} honest-but-curious cloud server, generating a new update key does not solve the problem, because the cloud server knows the just generated update key and thus it can update the revoked decryption keys.
In any case, the honest-but-curious cloud server and the colluding consumers cannot combine somehow the update keys and decryption keys to decrypt some data that singularly the colluding consumers cannot decrypt.
In Section \ref{sec:proof} we will capture this adversary model with the Game 2, and we will provide a formal proof that SEA-BREW is resistant against it.

\subsection{Scheme Definition}
\label{mathintro}
Our system makes use of a set of \emph{cryptographic primitives} (from now on, simply primitives), which are the following ones.
\newline\newline
$(\mathit{MK}, \mathit{EK}) = \mathbf{Setup}(\kappa)$: This primitive initializes the cryptographic scheme. 
It takes a security parameter $\kappa$ as input, and outputs a \emph{master key} $\mathit{MK}$ and an associated \emph{encryption key} $\mathit{EK}$. 
\newline\newline
$\mathit{CP} = \mathbf{Encrypt}(M, \mathcal{P}, \mathit{EK})$: This primitive encrypts a plaintext $M$ under the policy $\mathcal{P}$. 
It takes as input the message $M$, the encryption key $\mathit{EK}$, and the policy $\mathcal{P}$. 
It outputs the ciphertext $\mathit{CP}$.
\newline\newline
$\mathit{DK} = \mathbf{KeyGen}(\gamma, \mathit{MK})$: This primitive generates a decryption key. 
It takes as input a set of attributes $\gamma$ which describes the consumer, and the master key $\mathit{MK}$.
It outputs a decryption key $\mathit{DK}$, which is composed of two fields for each attribute in $\gamma$, plus a field called $\mathit{D}$, useful to update such a key.
\newline\newline
$M = \mathbf{Decrypt}(\mathit{CP}, \mathit{DK})$: This primitive decrypts a ciphertext $\mathit{CP}$. 
It takes the ciphertext $\mathit{CP}$ and the consumer's decryption key $\mathit{DK}$ as input, and outputs the message $M$ if decryption is successful, $\perp$ otherwise.
The decryption is successful if and only if $\gamma$ satisfies $\mathcal{P}$, which is embedded in $\mathit{CP}$.
\newline\newline
The following primitives use symbols with a superscript number to indicate the version of the associated quantity.
For example, $\mathit{MK^{(i)}}$ indicates the $i$-th version of the master key, $\mathit{DK^{(i)}}$ indicates the $i$-th version of a given decryption key, etc.
\newline\newline
$(\mathit{MK}^{(i+1)}, U^{(i+1)}) = \mathbf{UpdateMK}(\mathit{MK}^{(i)})$:
This primitive updates the master key from a version $i$ to the version $i+1$ after a key revocation. 
It takes as input the old master key $\mathit{MK}^{(i)}$, and it outputs an updated master key $\mathit{MK}^{(i+1)}$, and the $(i+1)$-th version of the update key $U^{(i+1)}$.
Such an update key is composed of the quantities $U_{EK}^{(i+1)}$, $U_{DK}^{(i+1)}$, $U_{CP}^{(i+1)}$, which will be used after a key revocation respectively to update the encryption key, to update the decryption keys, and to re-encrypt the ciphertexts.
\newline\newline
$\mathit{EK}^{(n)} = \mathbf{UpdateEK}(\mathit{EK}^{(i)}, U^{(n)}_{\mathit{EK}})$: 
This primitive updates an encryption key from a version $i$ to the latest version $n$, with $n>i$, after a key revocation.
The primitive takes as input the old encryption key $\mathit{EK}^{(i)}$ and $U^{(n)}_{\mathit{EK}}$, and it outputs the updated encryption key $\mathit{EK}^{(n)}$.
\newline\newline
$D^{(n)} = \mathbf{UpdateDK}(D^{(i)}, U^{(i)}_{DK}, U^{(i+1)}_{DK}, \dots, U^{(n)}_{DK})$: 
This primitive updates a decryption key from a version $i$ to the latest version $n$, with $n>i$, after a key revocation. 
What is actually updated is not the whole decryption key, but only a particular field $D$ inside the decryption key.
This allows the cloud server to execute the primitive without knowing the whole decryption key, but only $D$ which alone is useless for decrypting anything.
The primitive takes as input the old field $D^{(i)}$ and $U^{(i)}_{DK}, U^{(i+1)}_{DK}, \dots, U^{(n)}_{DK}$, and it outputs the updated field $D^{(n)}$.
\newline\newline
$\mathit{CP}^{(n)} = \mathbf{UpdateCP}(\mathit{CP}^{(i)}, U^{(i)}_{CP}, U^{(i+1)}_{CP}, \dots, U^{(n)}_{CP})$: This primitive updates a ciphertext from a version $i$ to the latest version $n$, with $n>i$, after a key revocation. 
The cloud server executes this primitive to perform proxy re-encryption on ciphertexts.
The primitive takes as input the old ciphertext $\mathit{CP}^{(i)}$, and $U^{(i)}_{CP}, U^{(i+1)}_{CP}, \dots, U^{(n)}_{CP}$. 
It outputs the updated ciphertext $\mathit{CP}^{(n)}$.
\newline
\newline
The concrete construction of these primitives will be described in detail in Section \ref{math}.
Moreover, SEA-BREW also needs a symmetric key encryption (e.g., AES, 3DES, \dots) scheme and a digital signature scheme (e.g., RSA, DSA, ECDSA, \dots). 
However, those will not be covered in this paper since such a choice does not affect the behavior of our system.

\subsection{Security Definition}
We state that SEA-BREW is secure against an adaptive chosen plaintext attack (IND-CPA) if no probabilistic polynomial-time (PPT) adversary $\mathcal{A}$ has a non-negligible advantage against the challenger in the following game, denoted as Game 1.
Note that IND-CPA security is not enough in the presence of an active adversary, however a stronger adaptive IND-CCA security assurance can be obtained in the random oracle model by means of the simple Fujisaki-Okamoto transformation \cite{fujisaki1999secure}, which only requires few additional hash computations in the $\mathrm{Encrypt}$ and the $\mathrm{Decrypt}$ primitives.

\paragraph{Setup}
The challenger runs the $\mathrm{Setup}$ primitive and generates $\mathit{EK^{(0)}}$, and sends it to the adversary.

\paragraph{Phase 1}
The adversary may issue queries for:
\begin{itemize}
    \item \textit{encryption key update}: the challenger runs the primitive $\mathrm{UpdateMK}$. The challenger sends the updated encryption key to the adversary.
    \item \textit{generate decryption key}: the challenger runs the primitive $\mathrm{KeyGen}$ using as input an attribute set provided by the adversary. Then, the challenger sends the generated decryption key to the adversary. 
    \item \textit{decryption key update}: the challenger runs the primitive $\mathrm{UpdateDK}$ using as input a decryption key provided by the adversary. Then, the challenger sends the updated decryption key to the adversary.
    \item \textit{ciphertext update}: the challenger runs the primitive $\mathrm{UpdateCP}$ using as input a ciphertext provided by the adversary. Then, the challenger sends the ciphertext updated to the last version to the adversary.
\end{itemize}

\paragraph{Challenge} 
The adversary submits two equal length messages $m_0$ and $m_1$ and a challenge policy $\mathcal{P}^*$, which is not satisfied by any attribute set queried as \textit{generate decryption key} during Phase 1.
The challenger flips a fair coin and assigns the outcome to $b$: $b\leftarrow\{0,1\}$. 
Then, the challenger runs the $\mathrm{Encrypt}$ primitive encrypting $m_b$ under the challenge policy $\mathcal{P}^*$ using $\mathit{EK^{(n)}}$ and sends the ciphertext $\mathit{CP}^*$ to the adversary. The symbol $n$ is the last version of the master key, i.e., the number of time the adversary queried for an encryption key update.

\paragraph{Phase 2}
Phase 1 is repeated.
However the adversary cannot issue queries for \textit{generate decryption key} whose attribute set $\gamma$ satisfies the challenge policy $\mathcal{P}^*$.

\paragraph{Guess}
The adversary outputs a guess $b^\prime$ of $b$. 
The advantage of an adversary $\mathcal{A}$ in Game 1 is defined as $\mathrm{Pr}[b^\prime=b]-\frac{1}{2}$.
\newline
\newline
We prove SEA-BREW to be secure in Section \ref{sec:proof}.

\section{SEA-BREW Procedures}
\label{sec:procedures}
In the following, we describe the procedures that our system performs.

\subsection{System Initialization}
The system initialization procedure is executed only once, to start the system, and it consists in the following steps. \newline
\textbf{Step 1.} The key authority runs the $\mathrm{Setup}$ primitive, thus obtaining the first version of the master key ($\mathit{MK}^{(0)}$) and the first version of the encryption key ($\mathit{EK}^{(0)}$).
We indicate with $\mathit{v_{MK}}$ (\emph{master key version}) the current version of the master key. 
The key authority initializes the master key version to $\mathit{v_{MK}}=0$, and it sends the encryption key and the master key version to the cloud server with a signed message. \newline
\textbf{Step 2.} The cloud server, in turn, sends the encryption key and the master key version to the WSAN gateway with a signed message. \newline
\textbf{Step 3.} The WSAN gateway generates the broadcast public key (see Figure \ref{fig:sm}) for the broadcast encryption scheme.

\subsection{Producer Join}
The consumer join procedure is executed whenever a new producer joins the system. 
We assume that the producer has already pre-installed its own pair of asymmetric keys that it will use for digital signatures.
Alternatively the producer can create such a pair at the first boot.
We call \emph{signature verification key} ($\mathit{SK}$, see Figure \ref{fig:sm}) the public key of such a pair. 
The procedure consists in the following steps. \newline
\textbf{Step 1.} The producer sends the signature verification key to the key authority in some authenticated fashion.
The mechanism by which this is done falls outside the scope of the paper.
For example, in case the producer is a sensor, the human operator who is physically deploying the sensor can leverage a pre-shared password with the key authority. \newline
\textbf{Step 2.} The key authority assigns a new producer identifier to the producer, and it sends such an identifier and the encryption key to the producer with a signed message.
The encryption key embeds an \emph{encryption key version} ($\mathit{v_{EK}}$), which represents the current version of the encryption key locally maintained by the producer.
Initially, the encryption key version is equal to the master key version ($\mathit{v_{EK}} = \mathit{v_{MK}}$). \newline
\textbf{Step 3.} The key authority also sends the producer's identifier, signature verification key and encryption key version to the cloud server with a signed message.
The cloud server adds a tuple $\langle \mathit{PID}, \mathit{SK}, \mathit{v_{EK}}\rangle$ to a locally maintained \emph{Producer Table} (PT, see Figure \ref{fig:sm}).
Each tuple in the PT represents a producer in the system.

If the producer is remote, then the procedure ends here.
Otherwise, if the producer is inside the WSAN, then the following additional steps are performed. \newline
\textbf{Step 4.} The key authority sends the producer identifier and the signature verification key to the WSAN gateway with a signed message.
The WSAN gateway adds a tuple $\langle \mathit{PID}, \mathit{SK}\rangle$ to a locally maintained \emph{WSAN Signature Table} (see Figure \ref{fig:sm}).
Each tuple in the WSAN Signature Table represents a producer in the WSAN.
Through this table, both the gateway and the consumers are able to authenticate data and messages generated by the producers in the WSAN. \newline
\textbf{Step 5.} The WSAN gateway finally broadcasts the signed message received from the key authority to all the WSAN consumers.
The WSAN consumers add the same tuple $\langle \mathit{PID}, \mathit{SK}\rangle$ to a locally maintained copy of the WSAN Signature Table.

\subsection{Consumer Join}
The consumer join procedure is executed whenever a new consumer, described by a given attribute set, joins the system. 
We assume that the consumer has already pre-installed its own pair of asymmetric keys that it will use for asymmetric encryption.
Alternatively the consumer can create such a pair at the first boot.
We call \emph{key distribution key} ($\mathit{KDK}$, see Figure \ref{fig:sm}) the public key of such a pair.
The procedure consists in the following steps. \newline
\textbf{Step 1.} The consumer sends the key distribution key to the key authority in some authenticated fashion. 
Again, the mechanism by which this is done falls outside the scope of the paper. \newline
\textbf{Step 2.} The key authority assigns a new consumer identifier to the consumer, and it generates a decryption key with the $\mathit{KeyGen}$ primitive, according to the consumer's attribute set. 
The key authority sends the consumer identifier and the decryption key to the consumer with a signed message, encrypted with the consumer's key distribution key. \newline
\textbf{Step 4.} The key authority sends the consumer identifier and the field $D$ of the decryption key to the cloud server with a signed message.
The cloud server initializes a \emph{decryption key version} ($\mathit{v_{DK}}$), which represents the current version of the consumer's decryption key, to the value of the master key version.
The cloud server adds a tuple $\langle \mathit{CID}, D, \mathit{v_{DK}}\rangle$ to a locally maintained \emph{Consumer Table} (CT, see Figure \ref{fig:sm}).
Each tuple in the CT represents a consumer in the system.

If the consumer is remote, then the procedure ends here.
Otherwise, if the consumer is a WSAN consumer, then the following additional steps are performed. \newline
\textbf{Step 5.} The key authority sends the consumer identifier and the key distribution key to the WSAN gateway with a signed message. \newline
\textbf{Step 6.} The WSAN gateway sends the WSAN Signature Table to the consumer with a signed message, along with the broadcast public key and the consumer's broadcast private key which is appropriately encrypted with the consumer's key distribution key.
Finally, the WSAN gateway adds a tuple $\langle \mathit{CID}, \mathit{KDK}\rangle$ to a locally maintained \emph{WSAN Consumer Table}.

\subsection{Data Upload by Remote Producers}
\label{daup}
The data upload procedure is executed whenever a producer wants to upload data to the cloud server.
Remote producers and WSAN producers perform two different procedures to upload a piece of information to the cloud server.
We explain them separately.
The data upload procedure by remote producers consists in the following steps. \newline
\textbf{Step 1.} Let $\mathcal{P}$ be the access policy that has to be enforced over the data. 
The remote producer encrypts the data under such a policy using the $\mathrm{Encrypt}$ primitive.
The resulting ciphertext has the same version number of the producer's locally maintained encryption key ($\mathit{v_{CP}}=\mathit{v_{EK}}$). \newline
\textbf{Step 2.} The producer securely deletes the original data.
Then it signs and uploads the ciphertext to the cloud server, along with its producer identifier. \newline
\textbf{Step 3.} The cloud server verifies the signature, and then it stores the ciphertext.
\newline
Finally, if the ciphertext version is older than the master key version, the cloud server executes the remote producer update procedure (see Section \ref{produp}).

\subsection{Data Upload by WSAN Producers}
\begin{figure}[t]
	\centering
	\includegraphics{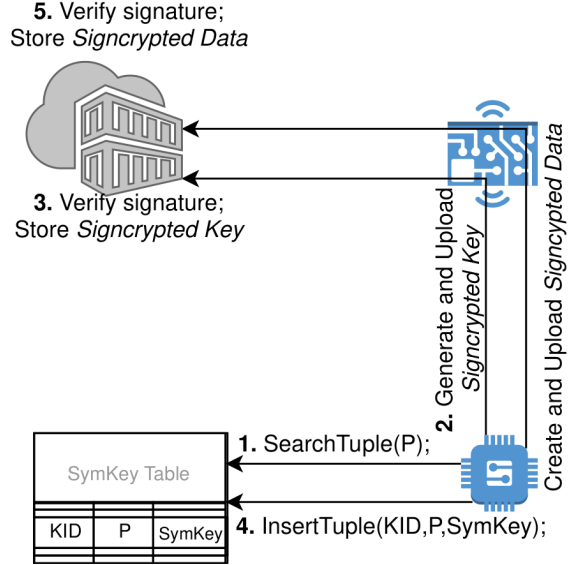}
	\caption{
	Data upload by WSAN producers procedure. 
	}
	\label{fig:daup}
\end{figure}
SEA-BREW aims at saving bandwidth in the WSAN also during data upload.
However, encrypting data directly with the $\mathit{Encrypt}$ primitive introduces a lot of overhead in terms of data size, as it happens in the typical ABE scheme.
Therefore, we want to obtain the access control mechanism provided by the $\mathit{Encrypt}$ primitive, and at the same time producing the small ciphertexts typical of symmetric-key encryption.
Broadly speaking, we achieve this by encrypting a symmetric key using the $\mathit{Encrypt}$ primitive, and then using such a symmetric key to encrypt all the data that must be accessible with the same access policy.
To do this, each WSAN producer maintains a \emph{SymKey Table} (see Figure \ref{fig:sm}), which associates policies $\mathcal{P}$ to symmetric keys $\mathit{SymKey}$.
More specifically, the SymKey Table is composed of tuples in the form $\langle KID,\mathcal{P},\mathit{SymKey}\rangle$, where $KID$ is the \emph{symmetric key identifier} of $SymKey$.
The symmetric key identifier uniquely identifies a symmetric key in the whole system.
The data upload procedure by WSAN producers consists in the following steps (Figure \ref{fig:daup}). \newline
\textbf{Step 1.} Let $\mathcal{P}$ be the access policy that has to be enforced over the data.
The producer searches for a tuple inside its SymKey Table associated with the policy.
If such a tuple already exists, then the producer jumps directly to Step 4, otherwise it creates it by continuing to Step 2. \newline
\textbf{Step 2.} The producer randomly generates a symmetric key and a symmetric key identifier.
The symmetric key identifier must be represented on a sufficient number of bits to make the probability that two producers choose the same identifier for two different symmetric keys negligible.
The producer then encrypts the symmmetric key under the policy using the $\mathrm{Encrypt}$ primitive, and it signs the resulting ciphertext together with the key identifier.
The result is the \emph{signcrypted key}.
The producer uploads the signcrypted key and its producer identifier to the cloud server.
\newline
\textbf{Step 3.} The cloud server verifies the signature, and then it stores the signcrypted key in the same way it stores ordinary encrypted data produced by remote producers. \newline
\textbf{Step 4.} The producer inserts (or retrieves, if steps 2 and 3 have not been executed) the tuple $\langle KID,\mathcal{P},\mathrm{SymKey}\rangle$ into (from) its SymKey Table, and it encrypts the data using the symmetric key associated to the policy. 
Then, the producer signs the resulting ciphertext together with the symmetric key identifier.
The result is the \emph{signcrypted data}.
The producer uploads the signcrypted data and its producer identifier to the cloud server, and it securely deletes the original data.\newline
\textbf{Step 5.} The cloud server verifies the signature, and then it stores the signcrypted data.

\subsection{Data Download}
\label{dado}
The data download procedure is executed whenever a consumer wants to download data from the cloud server.
Consumers perform two different procedures to download a piece of information from the cloud server, depending whether such piece of information has been produced by a remote producer or by a WSAN producer.
We explain them separately.
The download procedure of data produced by remote producers consists in the following steps. \newline
\textbf{Step 1.} The consumer sends a data request along with its consumer identifier to the cloud server. \newline
\textbf{Step 2.} The cloud server checks in the CT whether the decryption key version of the consumer is older than the master key version and, if so, it updates the decryption key by executing the remote consumer update procedure (see after).
The cloud server identifies the requested ciphertext and checks whether its version is older than the master key version.
If so, the cloud server updates the ciphertext by executing the $\mathrm{UpdateCP}$ primitive (see Section \ref{math}). \newline
\textbf{Step 3.} The cloud server signs and sends the requested data to the consumer. \newline
\textbf{Step 4.} The consumer verifies the server signature over the received message.
Then, it executes the $\mathrm{Decrypt}$ primitive using its decryption key.

Now consider the case in which a consumer requests a data produced by a WSAN producer.
\begin{figure}[t]
	\centering
	\includegraphics{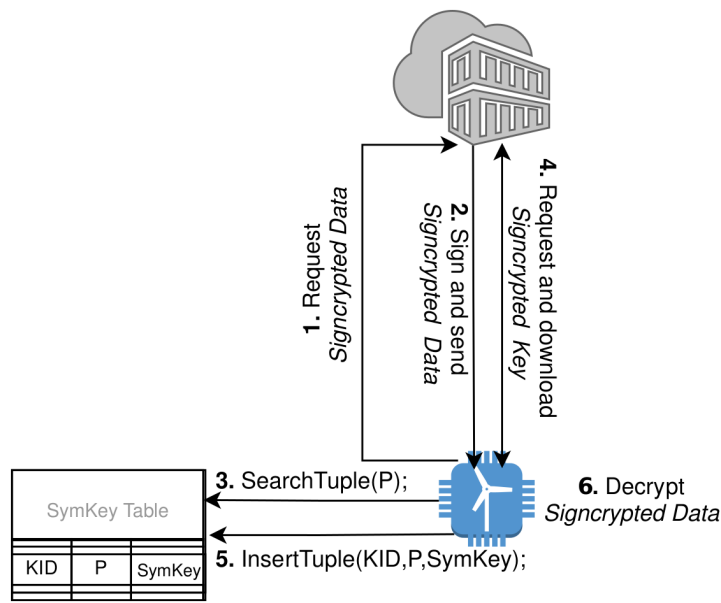}
	\caption{
	Download signcrypted data procedure. 
	}
	\label{fig:dado}
\end{figure}
Each consumer maintains a SymKey Table (see Figure \ref{fig:sm}), which associates policies $\mathcal{P}$ to symmetric keys $\mathit{SymKey}$.
The download procedure of data produced by WSAN producers consists in the following steps (Figure \ref{fig:dado}). \newline
\textbf{Step 1.} The consumer sends a data request along with its consumer identifier to the cloud server. \newline
\textbf{Step 2.} The cloud server signs and sends the requested signcrypted data to the consumer. \newline
\textbf{Step 3.} The consumer searches for a tuple with the same key identifier as the one contained in the received signcrypted data inside its SymKey Table.
If such a tuple already exists, then the consumer jumps directly to Step 6, otherwise the consumer creates it by continuing to Step 4. \newline
\textbf{Step 4.} The consumer performs a data download procedure, requesting and obtaining the signcrypted key associated to the received symmetric key identifier. \newline
\textbf{Step 5.} The consumer decrypts the signcrypted key thus obtaining the symmetric key, and it adds the tuple $\langle KID,\mathcal{P},\mathrm{SymKey}\rangle$ to its SymKey Table.\newline
\textbf{Step 6.} The consumer decrypts the signcrypted data with the symmetric key.

\subsection{Direct Data Exchange}
The direct data exchange procedure is executed whenever a producer wants to transmit data to one or more consumers in a low-latency fashion inside the WSAN.
To obtain a low latency the producer broadcasts the data directly to the authorized consumers in an encrypted form, instead of uploading such data to the cloud server.
Furthermore, to save WSAN bandwidth we want that the data exchanged is encrypted with symmetric-key encryption, under the form of signcrypted data as it happens for data uploaded by WSAN producers.
To ease the reading we assume that the producer has already a tuple associated to the policy it wants to apply.
Otherwise the producer should previously perform a data upload procedure to the cloud in which it uploads the signcrypted key it will use.

The procedure consists in the following steps. \newline
\textbf{Step 1.} Let $\mathcal{P}$ be the access policy that has to be enforced over the data.
The producer retrieves the symmetric key associated to such policy inside its SymKey Table.
The producer encrypts the data with such a symmetric key, and signs it together with the symmetric key identifier.
It thus obtains the signcrypted data. \newline
\textbf{Step 2.} The producer broadcasts the signcrypted data in the WSAN, and securely deletes the original data. \newline
\textbf{Step 3.} Perform Steps 3-6 of the download procedure of data produced by WSAN producers.

\subsection{Producer Leave}
The producer leave procedure is executed whenever one or more producers leave the system.
This happens in case that producers are dismissed from the system, or the private keys that they use for signatures are compromised.
In all these cases, the private keys of the leaving producers must be revoked, so that data signed with such keys is no longer accepted by the cloud server.
The procedure consists in the following steps. \newline
\textbf{Step 1.} The key authority communicates to the cloud server the identifiers of the leaving producers with a signed message. \newline
\textbf{Step 2.} The cloud server removes the tuples associated to such identifiers from the PT.

If at least one leaving producer was a WSAN producer, the following additional steps are performed. \newline
\textbf{Step 3.} The key authority communicates the identifiers of the leaving WSAN producers to the WSAN gateway with a signed message. \newline
\textbf{Step 4.} The WSAN gateway removes the tuples associated to such identifiers from the WSAN Signature Table, and it broadcasts the signed message received by the key authority to all the WSAN consumers. \newline
\textbf{Step 5.} The WSAN consumers remove the tuples associated to such identifiers from their locally maintained copy of the WSAN Signature Table.

\subsection{Consumer Leave}
\begin{figure}[t]
	\centering
	\includegraphics[scale=0.8]{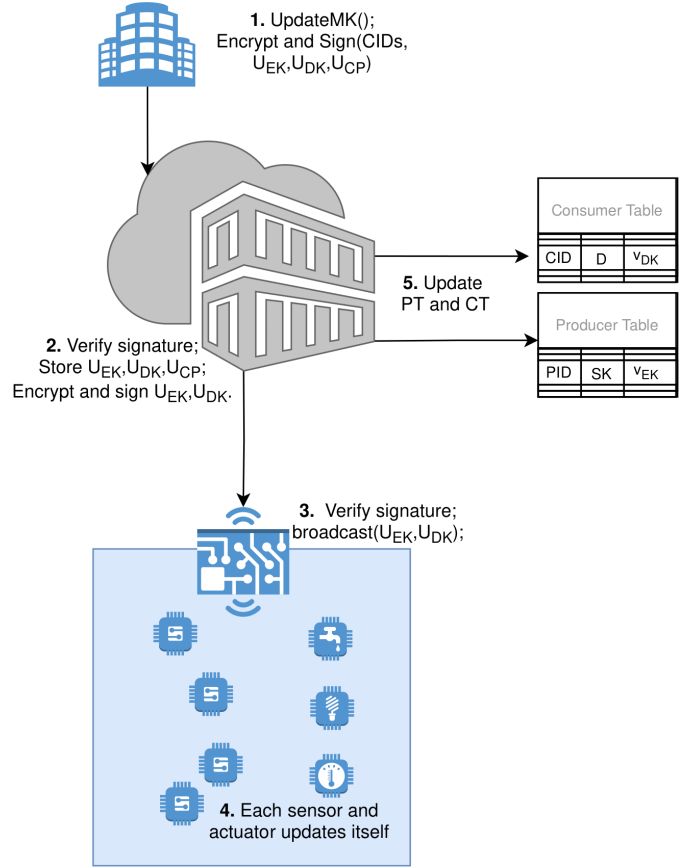}
	\caption{
	Consumer leave procedure. 
	}
	\label{fig:cole}
\end{figure}
The consumer leave procedure is executed whenever one or more consumers leave the system, as depicted in figure \ref{fig:cole}.
This happens in case that consumers are dismissed from the system, or their keys are compromised.
In all these cases, the decryption keys of the leaving consumers must be revoked, in such a way that they cannot decrypt data anymore.
The procedure consists in the following steps. \newline
\textbf{Step 1.} The key authority increases the master key version, and it executes the $\mathrm{UpdateMK}$ primitive on the old master key, thus obtaining the new master key and the quantities $U^{(\mathit{v_{MK}})}_{EK}$, $U^{(\mathit{v_{MK}})}_{DK}$, and $U^{(\mathit{v_{MK}})}_{CP}$.
Then, the key authority sends the identifiers of the leaving consumers and the quantities $U^{(\mathit{v_{MK}})}_{EK}$, $U^{(\mathit{v_{MK}})}_{DK}$, and $U^{(\mathit{v_{MK}})}_{CP}$ to the cloud server with a signed message, encrypted with the cloud server's public key. \newline
\textbf{Step 2.} The cloud server verifies the signature, decrypts the message, retrieves the consumer identifier from the message, and removes the tuples associated to those identifiers from the CT.
Note that the cloud server could now re-encrypt all the ciphertexts, by using the quantity $U^{(\mathit{v_{MK}})}_{CP}$ just received.
However, the re-encryption of each ciphertext is deferred to the time at which a consumer requests it (Lazy PRE). 
Then, the cloud server signs and encrypts $U_{EK}^{(\mathit{v_{MK}})}$ and $U_{DK}^{(\mathit{v_{MK}})}$ with asymmetric encryption, and it sends them to the gateway. \newline
\textbf{Step 3.} The gateway broadcasts the quantity $U_{EK}^{(\mathit{v_{MK}})}$ and $U_{DK}^{(\mathit{v_{MK}})}$ over the local low-bitrate WSAN, so that all the producers and consumers that belong to it can immediately update their encryption key and decryption key, respectively.
To do this the gateway sends a single broadcast message, composed as follows.
The gateway encrypts the $U_{DK}^{(\mathit{v_{MK}})}$ quantity with the broadcast public key, in such a way that all the WSAN consumers except the leaving ones can decrypt it.
This allows the gateway to share said quantity only with the WSAN consumers, excluding the compromised ones if there are any.
The gateway then signs the concatenation of the quantity $U_{EK}^{(\mathit{v_{MK}})}$, and the quantity $U_{DK}^{(\mathit{v_{MK}})}$ (encrypted), and broadcasts said message over the WSAN. \newline
\textbf{Step 4.} Each producer updates its encryption key upon receiving the broadcast message;
each consumer then decrypts the received message using its broadcast private key $d_{CID}$, and executes the $\mathrm{UpdateDK}$ primitive using its old decryption key and the just received $U_{DK}^{(\mathit{v_{MK}})}$.
The WSAN producers and the consumers delete their SymKey Tables. \newline
\textbf{Step 5.} The cloud server updates inside the PT the versions of the encryption keys of all the WSAN producers, and inside the CT the versions of the decryption keys of all the WSAN consumers.

Note that SEA-BREW updates all the devices inside the low-bitrate WSAN with a single $\mathcal{O}(1)$-sized broadcast message (Step 3).
This makes SEA-BREW highly scalable in the number and size of messages necessary to manage decryption keys.
Note also that, regarding remote consumers and remote producers, the computational load of the consumer leave procedure is entirely delegated to the cloud server, leaving the producers and consumers free of heavy computation. 
This enables SEA-BREW to run on a broader class of sensors and actuators.

\subsection{Remote Producer Update}
\label{produp}
The producer update procedure is executed by the data upload procedure by remote producers (see Section \ref{daup}), and it consists in the following steps.
\textbf{Step 1.} The cloud server signs and sends the last quantity $U_{EK}$ received from the key authority to the remote producer that must be updated. \newline
\textbf{Step 2.} The producer verifies the signature and retrieves $U_{EK}$.
Then, it executes the $\mathrm{UpdateEK}$ primitive using its encryption key and the received quantity $U_{EK}$ as parameters. \newline
\textbf{Step 3.} The cloud server updates the producer's encryption key version to $\mathit{v_{MK}}$ inside PT.

\subsection{Remote Consumer Update}
The consumer update procedure is executed as specified in the data download procedure (see Section \ref{dado}), and it consists in the following steps.\newline
\textbf{Step 1.} The cloud server executes the $\mathrm{UpdateDK}$ primitive using the consumer's decryption key and the last $(\mathit{v_{MK}} - \mathit{v_{DK}})$ quantities $U_{DK}$s received from the key authority.
The cloud server encrypts and signs the output of that primitive, $D^{(\mathit{v_{MK}})}$ using the consumer's key-encryption key, and sends it to the consumer. \newline
\textbf{Step 2.} The consumer verifies the signature and decrypts the message, thus obtaining back $D^{(\mathit{v_{MK}})}$.
Then, the consumer replaces the old field $D$ of its decryption key with the received quantity.\newline
\textbf{Step 3.} The cloud server updates the consumer's decryption key version to $\mathit{v_{MK}}$ inside CT.

\section{Concrete Construction}
\label{math}
We now explain in detail how the CP-ABE primitives previously introduced at the beginning of Section \ref{mathintro} are realized.
\newline
\newline
$(\mathit{MK}^{(0)}, \mathit{EK}^{(0)}) = \mathrm{Setup}(\kappa)$\newline
The $\mathrm{Setup}$ primitive is executed by the key authority.
This primitive computes:
\begin{equation}
\mathit{EK}^{(0)} = \{\mathds{G}_0, g, h = g^{\beta}, l=e(g,g)^\alpha,\mathit{v_{EK}}=0\};
\end{equation}
\begin{equation}
\mathit{MK}^{(0)} = \{\beta, g^\alpha,\mathit{v_{MK}}=0\},
\end{equation}
where $\mathds{G}_0$ is a multiplicative cyclic group of prime order $p$ with size $\kappa$, $g$ is the generator of $\mathds{G}_0$, $e : \mathds{G}_0 \times \mathds{G}_0 \rightarrow \mathds{G}_1$ is an efficiently-computable bilinear map with bi-linearity and non-degeneracy properties, and $\alpha,\beta\in\mathds{Z}_p$ are chosen at random.
\newline
\newline
$\mathit{CP} = \mathrm{Encrypt}(M, \mathcal{P}, \mathit{EK}^{(\mathit{v_{EK}})})$ \newline 
The $\mathrm{Encrypt}$ primitive is executed by a producer.
From now on, $\mathcal{P}$ is represented as a \emph{policy tree}, which is a labeled tree where the non-leaf nodes implement \emph{threshold-gate operators} whereas the leaf nodes are the attributes of the policy.
A threshold-gate operator is a Boolean operator of the type $k$-of-$n$, which evaluates to true iff at least $k$ (\emph{threshold value}) of the $n$ inputs are true.
Note that a $1$-of-$n$ threshold gate implements an OR operator, whereas an $n$-of-$n$ threshold gate implements an AND operator.
For each node $x$ belonging to the policy tree the primitive selects a polynomial $q_x$ of degree equal to its threshold value minus one ($d_x = k_x - 1$). 
The leaf nodes have threshold value $k_x=1$, so their polynomials have degree equal to $d_x=0$.
The polynomials are chosen in the following way, starting from the root node $R$. 
The primitive assigns arbitrarily an index to each node inside the policy tree. 
The index range varies from $1$ to $num$, where $num$ is the total number of the nodes. 
The function $\operatorname{index}(x)$ returns the index assigned to the node $x$.
Starting with the root node $R$ the primitive chooses a random $s\in Z_p$ and sets $q_R(0) = s$. 
Then, it randomly chooses $d_R$ other points of the polynomial $q_R$ to completely define it. 
Iteratively, the primitive sets $q_x(0) = q_{\operatorname{parent}(x)}(\operatorname{index}(x))$ for any other node $x$ and randomly chooses $d_x$ other points to completely define $q_x$, where $\operatorname{parent}(x)$ refers to the parent of the node $x$.
At the end, the ciphertext is computed as follows:
\begin{equation}
\begin{split}
\mathit{CP}=\{&\mathcal{P}, \tilde{C}=Me(g,g)^{\alpha s}, C = h^s, \mathit{v_{CP}}=\mathit{v_{EK}}\\
&\forall y \in Y: \hspace{13pt} C_y=g^{q_y(0)}, C_{y}'=H(\operatorname{att}(y))^{q_y(0)} \},
\end{split}
\end{equation}
where $Y$ is the set of leaf nodes of the policy tree.
The function $\operatorname{att}(x)$ is defined only if $x$ is a leaf node, and it denotes the attribute associated with the leaf.
$H$ is a hash function $H :\{0,1\}^* \rightarrow \mathds{G}_0$ that is modeled as a random oracle. 
The encryption key version $\mathit{v_{EK}}$ is assigned to the ciphertext version $\mathit{v_{CP}}$.
\newline
\newline
$\mathit{DK} = \mathrm{KeyGen}(\mathit{MK}^{(\mathit{v_{MK}})}, \gamma)$ \newline 
The $\mathrm{KeyGen}$ primitive is executed by the key authority.
This primitive randomly selects $r \in \mathds{Z}_p$, and $r_j \in \mathds{Z}_p$ for each attribute in $\gamma$.
It computes the decryption key $\mathit{DK}$ as:
\begin{equation}
\begin{split}
\mathit{DK} = \{& D=g^{\frac{(\alpha + r)}{\beta}}, \mathit{v_{DK}}=\mathit{v_{MK}}\\
&\forall j \in \gamma: \hspace{13pt} D_j=g^r\cdot H(j)^{r_j}, D'_j=g^{r_j}\}.
\end{split}
\end{equation}
\newline
\newline
$M = \mathrm{Decrypt}(\mathit{CP}, \mathit{DK})$ \newline
The $\mathrm{Decrypt}$ primitive is executed by a consumer.
This primitive executes the sub-function \textit{DecryptNode} on the root node. 
\textit{DecryptNode}$(\mathit{DK}, \mathit{CP}, x)$ takes as input the consumer's decryption key, the ciphertext and the node $x$.
If the node $x$ is a leaf node, let $i= \operatorname{att}(x)$ and define the function as follows.
If $i \in \gamma$, then:
\begin{equation}
\begin{split}
DecryptNode(\mathit{DK},\mathit{CP},&x) 	= \frac{e(D_i,C_x)}{e(D'_i,C'_x)}.
\end{split}
\end{equation}
Otherwise, if $i \notin \gamma$, then \textit{DecryptNode}$(\mathit{DK}, \mathit{CP}, x)=\perp$.
When $x$ is not a leaf node, the primitive proceeds as follows.
First of all, let $\Delta_{i,S}$ be the Lagrange coefficient for $i \in \mathds{Z}_p$ and let $S$ be an arbitrary set of element in $\mathds{Z}_p:\Delta_{i,S}(x)=\prod_{j\in S,j\neq i}\frac{x-j}{i-j}$.
Now, for all nodes $z$ that are children of $x$, it calls recursively itself and stores the result as $F_z$. 
Let $S_x$ be an arbitrary $k_x$-sized set of children $z$ such that $F_z \neq \perp \forall z \in S_x$. 
Then, the function computes:
\begin{equation}
\label{eq:fx}
F_x = \prod_{z\in S_z} F_z^{\Delta_{i,S'_x}(0)}= e(g,g)^{r\cdot q_x(0)}.
\end{equation}
where $i=\operatorname{index}(z)$, and $S_x = {\operatorname{index}(z):z\in S_x}$.
The $\mathrm{Decrypt}(\mathit{CP}, \mathit{DK})$ primitive first calls \textit{DecryptNode}$(\mathit{DK}, \mathit{CP}, R)$ where $R$ is the root of the policy tree extracted by $\mathcal{P}$ embedded in $\mathit{CP}$.
Basically, the sub-function navigates the policy tree embedded inside the ciphertext in a top-down manner and if $\gamma$ satisfies the policy tree it returns $A = e(g,g)^{rs}$.
Finally, the primitive computes:
\begin{equation}
M=\tilde{C}/(e(C,D)/A).
\label{cb}
\end{equation}
\newline
\newline
$(\mathit{MK}^{(\mathit{v_{MK}}+1)},U^{(\mathit{v_{MK}}+1)}) = \mathrm{UpdateMK}(\mathit{MK}^{(\mathit{v_{MK}})})$\newline 
The $\mathrm{UpdateMK}$ primitive is executed by the key authority.
This primitive increments $\mathit{v_{MK}}$ by one, chooses at random a new $\beta^{(\mathit{v_{MK}})} \in \mathds{Z}_p$, and computes: 
\begin{equation}
\begin{split}
U_{CP}^{(\mathit{v_{MK}})} &= \frac{\beta^{(\mathit{v_{MK}})}}{\beta^{(\mathit{v_{MK}-1})}} ;\\
U_{EK}^{(\mathit{v_{MK}})} &=  g^{\beta^{(\mathit{v_{MK}})}};\\
U_{DK}^{(\mathit{v_{MK}})} &=  \frac{\beta^{(\mathit{v_{MK}-1})}}{\beta^{(\mathit{v_{MK}})}} ;\\
U^{(\mathit{v_{MK}})} &=  \{U_{CP}^{(\mathit{v_{MK}})}, U_{EK}^{(\mathit{v_{MK}})}, U_{DK}^{(\mathit{v_{MK}})}\}.
\end{split}
\end{equation}
Then it updates the master key as:
\begin{equation}
\mathit{MK}^{(\mathit{v_{MK}})} = \{\beta^{(\mathit{v_{MK}})}, g^{\alpha},\mathit{v_{MK}}\}.
\end{equation}
In order to avoid ambiguities, we specify that the first ever update key is $U^{(1)}$ and not $U^{(0)}$ as the value $\mathit{v_{MK}}$ is incremented \emph{before} the creation of $U$.
The careful reader surely have noticed that $U_{CP}$ and $U_{DK}$ are reciprocal. 
In practice, we can use only one of these quantities and compute the other by inverting it. 
In this paper we chose to keep those quantity separated for the sake of clarity.
\newline
\newline
$EK^{(\mathit{v_{MK}})}=\mathrm{UpdateEK}(EK^{(\mathit{v_{EK}})},U_{EK}^{(\mathit{v_{MK}})})$ \newline 
The $\mathrm{UpdateEK}$ primitive is executed by the producers.
Regardless the input encryption key's version, this primitive takes as input only the last update key generated, namely $U_{EK}^{(\mathit{v_{MK}})}$.
The primitive substitutes the field $h$ inside the encryption key with the last update quantity, and updates the encryption key version to the latest master key version, thus obtaining:
\begin{equation}
\mathit{EK}^{(v_{MK})} = \{\mathds{G}_0, g, h = U_{EK}^{(\mathit{v_{MK}})}, l=e(g,g)^\alpha,\mathit{v_{EK}}=\mathit{v_{MK}}\}.
\end{equation}
\newline
\newline
$D^{(\mathit{v_{MK}})}=\mathrm{UpdateDK}(U_{DK}^{(\mathit{v_{DK}}+1)},\dots,U_{DK}^{(\mathit{v_{MK}})}, D^{(\mathit{v_{DK}})})$ \newline 
The $\mathrm{UpdateDK}$ primitive is executed by the cloud server and by the WSAN consumers.
The decryption key on input has been lastly updated with $U_{DK}^{(\mathit{v_{DK}})}$, and the overall latest update is $U_{DK}^{(\mathit{v_{MK}})}$, with, $\mathit{v_{MK}}>\mathit{v_{DK}}$. 
This primitive computes:
\begin{equation}
\label{eq:udk}
\begin{split}
U_{DK}'&=U_{DK}^{(\mathit{v_{DK}}+1)} \cdot \dots \cdot U_{DK}^{(\mathit{v_{MK}})};\\
D^{(\mathit{v_{MK}})} &= (D^{(\mathit{v_{DK}})})^{U_{DK}'}.
\end{split}
\end{equation}
\newline
\newline
$\mathit{CP}^{(\mathit{v_{MK}})}=\mathrm{UpdateCP}(\mathit{CP}^{(\mathit{v_{CP}})},U_{CP}^{(\mathit{v_{CP}}+1)},\dots,\mathit{U_{CP}}^{(\mathit{v_{MK}})})$ \newline
The $\mathrm{UpdateCP}$ primitive is executed by the cloud server.
The ciphertext on input has been lastly re-encrypted with $U_{CP}^{(\mathit{v_{CP}})}$, and the overall latest update is $U_{CP}^{(\mathit{v_{MK}})}$, with, $\mathit{v_{MK}}>\mathit{v_{CP}}$. 
This primitive computes the re-encryption quantity $U_{CP}'$ as the multiplication of all the version updates successive to the one in which the ciphertext has been lastly updated.
\begin{equation}
U_{CP}'=U_{CP}^{(\mathit{v_{CP}}+1)} \cdot \dots \cdot U_{CP}^{(\mathit{v_{MK}})}.
\end{equation}
Then, re-encryption is achieved with the following computation:
\begin{equation}
\label{eq:ucp}
C^{(\mathit{v_{MK}})}=(C^{(\mathit{v_{CP}})})^{U_{CP}'}.
\end{equation}
Finally, the primitive outputs the re-encrypted ciphertext $\mathit{CP'}$ as:
\begin{equation}
\begin{split}
\mathit{CP}^{(\mathit{v_{MK}})}&=\{\mathcal{P}, \tilde{C}, C^{(\mathit{v_{MK}})}, \mathit{v_{CP}}=\mathit{v_{MK}},\\
&\forall y \in Y: \hspace{13pt} C_y=g^{q_y(0)}, C_{y}'=H(\operatorname{att}(y))^{q_y(0)} \}.
\end{split}
\end{equation}

\subsection{Correctness.} 
In the following we show the correctness of SEA-BREW.

$\mathrm{Decrypt}$ equation (\ref{eq:fx}):
\begin{equation}
\begin{split}
F_x &= \prod_{z\in S_z} F_z^{\Delta_{i,S'_x}(0)}\\
&= \prod_{z\in S_z} (e(g,g)^{r\cdot q_z(0)})^{\Delta_{i,S'_x}(0)}\\
&= \prod_{z\in S_z} (e(g,g)^{r\cdot q_{\operatorname{parent}(z)}(\operatorname{index}(z))})^{\Delta_{i,S'_x}(0)}\\
&= \prod_{z\in S_z} e(g,g)^{r\cdot q_x(i) \cdot \Delta_{i,S'_x}(0)}\\
&= e(g,g)^{r\cdot q_x(0)}.
\end{split}
\end{equation}

$\mathrm{Decrypt}$ equation (\ref{cb}):
\begin{equation}
\begin {split}
\tilde{C}/(e(C,D)/A) 	&= \tilde{C}/(e(h^s,g^{\frac{\alpha + r}{\beta}})/e(g,g)^{rs})\\
&= Me(g,g)^{\alpha s}/ \left( e(g,g)^{\beta s \cdot \frac{\alpha + r}{\beta}}/e(g,g)^{rs} \right)\\
&= \frac{Me(g,g)^{\alpha s}}{e(g,g)^{\alpha s}} =M.
\end{split}
\end{equation}

$\mathrm{UpdateDK}$ equation (\ref{eq:udk}):
\begin{equation}
D^{(\mathit{v_{MK}})} = (D^{(\mathit{v_{DK}})})^{U_{DK}'} =g^{\frac{r + \alpha}{\beta^{(\mathit{v_{DK}})}}\cdot \frac{\beta^{(\mathit{v_{DK}})}}{\beta^{(\mathit{v_{MK}})}}}=g^{\frac{r + \alpha}{\beta^{(\mathit{v_{MK}})}}}.
\end{equation}

$\mathrm{UpdateCP}$ equation (\ref{eq:ucp}):
\begin{equation}
C^{(\mathit{v_{MK}})}=(C^{(\mathit{v_{CP}})})^{U_{CP}'}=g^{s\beta^{(\mathit{v_{CP}})}\cdot \frac{\beta^{(\mathit{v_{MK}})}}{\beta^{(\mathit{v_{CP}})}}}=g^{s\beta^{(\mathit{v_{MK}})} }.
\end{equation}

\section{Security Proofs}
\label{sec:proof}
In this section, we provide formal proofs of two security properties of our scheme, related to two adversary models described in Section \ref{sec:threat}.
Namely, we prove our scheme to be adaptively IND-CPA secure against a set of colluding consumers (Theorem \ref{thm:game1}), and against a honest-but-curious cloud server colluding with a set of consumers (Theorem \ref{thm:game2}).
\begin{theorem}
\label{thm:game1}
SEA-BREW is secure against an IND-CPA by a set of colluding consumers (Game 1), under the generic bilinear group model.
\end{theorem}

\begin{proof} 
Our objective is to show that SEA-BREW is not less secure than the CP-ABE scheme by Bethencourt et al. \cite{bethencourt2007ciphertext}, which is proved to be IND-CPA secure under the generic bilinear group model. 
To do this, we prove that if there is a PPT adversary $\mathcal{A}$ that can win Game 1 with non-negligible advantage $\epsilon$ against SEA-BREW, then we can build a PPT simulator $\mathcal{B}$ that can win the CP-ABE game described in \cite{bethencourt2007ciphertext} (henceforth, Game 0) against the scheme of Bethencourt et al. with the same advantage.
We will denote the challenger of Game 0 as $\mathcal{C}$.
We describe the simulator $\mathcal{B}$ in the following.

\paragraph{Setup}
In this phase $\mathcal{C}$ gives to $\mathcal{B}$ the public parameters $EK$ of Game 0, that will be exactly $\mathit{EK^{(0)}}$ in Game 1. 
In turn, $\mathcal{B}$ sends to $\mathcal{A}$ the encryption key $\mathit{EK^{(0)}}$ of Game 1.
\paragraph{Phase 1}
Let us denote with the symbol $n$ the latest version of the master key at any moment. 
In addition let us denote with the symbol $k$ a specific version of a key or a ciphertext lower than $n$, so that $k<n$ at any moment.
The query that an adversary can issue to the simulator are the following.
\begin{itemize}

    \item \textit{encryption key update}: $\mathcal{B}$ chooses $U_{DK}^{(n+1)}$ at random from $\mathds{Z}_p$.
    Then, $\mathcal{B}$ computes
    \begin{equation}
        h^{(n+1)}=(g^{\beta^{(n)}})^{\frac{1}{U_{DK}^{(n+1)}}},
    \end{equation}
    and sends $\mathit{EK^{(n+1)}}$ to $\mathcal{A}$.
    Finally, $\mathcal{B}$ increments $n$.
    Please note that $\mathcal{B}$ does not know $\beta^{(i)}, \forall i \in [0,n]$, but it does not need to.
    $\mathcal{B}$ needs to know only the relationship between any two consecutive versions, which are exactly: 
    \begin{equation}
        U_{DK}^{(i)} = \frac{\beta^{(i-1)}}{\beta^{(i)}}, \forall i \in [1,n]
    \end{equation}
    
    \item \textit{generate decryption key}: when $\mathcal{A}$ issues a query for $\mathit{DK_{j}^{(n)}}$ (i.e., a decryption key with a given attribute set $\gamma_j$, and latest version $n$) to $\mathcal{B}$, $\mathcal{B}$ in turn issues a query for $\mathit{DK_{j}}$ to $\mathcal{C}$, and receives $\mathit{DK_{j}^{(0)}}$. Then $\mathcal{B}$ upgrades such a key to the latest version $n$ executing the primitive $\mathrm{UpdateDK}$, using as input said key and $U_{DK}^{(i)}, \forall i \in [1,n]$. 
    Finally $\mathcal{B}$ sends to $\mathcal{A}$ the desired decryption key $\mathit{DK_{j}^{(n)}}$.
    
    \item \textit{decryption key update}: when $\mathcal{A}$ issues a query for upgrading an existing decryption key $\mathit{DK_{w}^{(k)}}$, $\mathcal{B}$ upgrades such a key to the last version $n$ executing the primitive $\mathrm{UpdateDK}$, using as input said key and $U_{DK}^{(i)}, \forall i \in [k,n]$. 
    Finally $\mathcal{B}$ sends to $\mathcal{A}$ the updated decryption key $\mathit{DK_{w}^{(n)}}$.
    
    \item \textit{ciphertext update}: when $\mathcal{A}$ issues a query for upgrading an existing ciphertext $\mathit{CP^{(k)}}$, $\mathcal{B}$ upgrades such a ciphertext to the latest version $n$ executing the primitive $\mathrm{UpdateCP}$, using as input said ciphertext and ${(U_{DK}^{(i)})}^{-1}, \forall i \in [k,n]$. 
    Finally $\mathcal{B}$ sends to $\mathcal{A}$ the updated ciphertext $\mathit{CP^{(n)}}$.

\end{itemize}

\paragraph{Challenge}

$\mathcal{A}$ submits two equal length messages $m_0$ and $m_1$ and a challenge policy $\mathcal{P}^*$ to $\mathcal{B}$, which in turn forwards them to $\mathcal{C}$.
$\mathcal{C}$ responds with $\mathit{CP}^*$ to $\mathcal{B}$, that will be exactly $\mathit{CP}^{*(0)}$ of Game 1. Then, $\mathcal{B}$ upgrades such a ciphertext to the latest version $n$ executing the primitive $\mathrm{UpdateCP}$, using as input said ciphertext and ${(U_{DK}^{(i)})}^{-1}, \forall i \in [1,n]$. Finally $\mathcal{B}$ sends to $\mathcal{A}$ the updated challenge ciphertext $\mathit{CP^{*(n)}}$.

\paragraph{Phase 2}
Phase 1 is repeated.

\paragraph{Guess}
$\mathcal{A}$ outputs $b^\prime$ to $\mathcal{B}$, which forwards it to $\mathcal{C}$.

Since a correct guess in Game 1 is also a correct guess in Game 0 and vice versa, then the advantage of the adversary $\mathcal{A}$ in Game 1 is equal to that of the adversary $\mathcal{B}$ in Game 0.
Namely, such an advantage is $\epsilon = \mathcal{O}(q^2/p)$, where $q$ is a bound on the total number of group elements received by the $\mathcal{A}$'s queries performed in Phase 1 and Phase 2, which is negligible with the security parameter $\kappa$.

    Please note that, in the encryption key update query, the adversary $\mathcal{A}$ cannot distinguish an $U_{DK}^{(i)}$ provided by $\mathcal{B}$ from one provided by the real scheme.
    Indeed, even if the generation of such a quantity is different, its probability distribution is uniform in $\mathds{Z}_p$ as in the real scheme. 
    This allows the simulator $\mathcal{B}$ to answer to all the other queries in Phase 1 and Phase 2 in a way that it is indistinguishable from the real scheme.
This concludes our proof. 
\end{proof}

We now consider a honest-but-curious cloud server colluding with a set of consumers.
We state that a scheme is secure against an IND-CPA by a honest-but-curious cloud server colluding with a set of consumers if no PPT adversary $\mathcal{A}$ has a non-negligible advantage against the challenger in the following game, denoted as Game 2. 
Game 2 is the same as Game 1 except that: (i) for every \textit{encryption key update} query in Phase 1 and Phase 2 the adversary is given also the update quantities $U_{DK}^{(i)}, \forall i \in [1,n]$; and (ii) during Phase 1 and Phase 2 the adversary can issue the following new type of query.
\begin{itemize}
    \item \textit{generate decryption key's $D$ field}: the challenger runs the primitive $\mathrm{KeyGen}$ using as input an attribute set provided by the adversary. Then, the challenger sends the field $D$ of generated decryption key to the adversary.
\end{itemize}
Note that differently from the \textit{generate decryption key} query, when issuing a \textit{generate decryption key's $D$ field} query the adversary is allowed to submit an attribute set that \textit{satisfies} the challenge policy $\mathcal{P}^*$.

\begin{theorem}
\label{thm:game2}
SEA-BREW is secure against an IND-CPA by a honest-but-curious cloud server colluding with a set of consumers (Game 2), under the generic bilinear group model.
\end{theorem}

\begin{proof}
We prove that if there is a PPT adversary $\mathcal{A}$ that can win Game 2 with non-negligible advantage $\epsilon$ against SEA-BREW, then we can build a PPT simulator $\mathcal{B}$ that can win Game 1 against SEA-BREW with the same advantage.
We can modify the simulator $\mathcal{B}$ used in the proof of Theorem \ref{thm:game1} to prove this theorem. 
In the Phase 1 and Phase 2, $\mathcal{B}$ additionally gives to $\mathcal{A}$ the update quantities $U_{DK}^{(i)}, \forall i \in [1,n]$, which $\mathcal{B}$ creates at each \textit{encryption key update} query.
During Phase 1 and Phase 2, when $\mathcal{A}$ issues a \textit{generate decryption key's $D$ field} query, $\mathcal{B}$ treats it in the same way of a \textit{generate decryption key} query with an empty attribute set $\gamma = \{\emptyset\}$.
Note indeed that a decryption key component $\mathit{D}_{\gamma_j}$ is indistinguishable from a \emph{complete} decryption key with \emph{no attributes}.
Hence, we can say that the advantage of $\mathcal{A}$ in Game 2 is the same as that of $\mathcal{B}$ in Game 0. 
Namely, such an advantage is $\epsilon = \mathcal{O}(q^2/p)$, which is negligible with the security parameter $\kappa$.
\end{proof}

\section{Performance Evaluation}
\label{sec:simulations}

In this section we analytically estimate the performances of SEA-BREW compared to: (i) the Bethencourt et al.'s scheme \cite{bethencourt2007ciphertext} provided with a simple key revocation mechanism, denoted as ``BSW-KU'' (Bethencourt-Sahai-Waters with Key Update); and (ii) Yu et al. scheme \cite{yu2010achieving}, denoted as ``YWRL'' (Yu-Wang-Ren-Lou).
We considered these two schemes for different reasons.
BSW-KU represents the simplest revocation method that can be built upon the ``classic'' CP-ABE scheme of Bethencourt et al.
Thus the performance of this revocation method constitutes the baseline reference for a generic revocable CP-ABE scheme.
On the other hand, YWRL represents a KP-ABE counterpart of SEA-BREW, since it natively supports an immediate indirect key revocation, and a Lazy PRE mechanism.

The revocation mechanism of BSW-KU works as follows.
The producer leave procedure works in the same way as SEA-BREW: the WSAN gateway simply broadcasts a signed message containing the producer identifier to all the WSAN consumers, which remove the tuples associated to such an identifier from their locally maintained copy of the WSAN Signature Table.
The consumer leave procedure requires the WSAN gateway to send a signed broadcast message containing the new encryption key to all the WSAN producers, and in addition an encrypted and signed message containing a new decryption key to each WSAN consumer.
This procedure results in $\mathcal{O}(n)$ point-to-point messages where $n$ is the number of WSAN consumers.
In contrast, SEA-BREW is able to perform both a consumer leave procedure by sending a single $\mathcal{O}(1)$-sized signed broadcast message over the WSAN.

\subsection{WSAN Traffic Overhead}
In this section we analytically estimate the traffic overhead that the key revocation mechanism of SEA-BREW generates in the WSAN, compared to the simple key revocation mechanism of BSW-KU.
In both SEA-BREW and BSW-KU schemes, for implementing $\mathds{G}_0$, $\mathds{G}_1$, and the bilinear pairing we consider a supersingular elliptic curve with embedding degree $k=2$ defined over a finite field of 512 bits.
For the signatures of the unicast and broadcast messages we consider a 160-bit ECDSA scheme.
Moreover, for the selective broadcast encryption used in the SEA-BREW scheme we consider the Boneh et al. scheme \cite{boneh2005collusion} with the same supersingular elliptic curve as above.
This gives to both schemes an overall security level of 80 bits.
We assume that, in both SEA-BREW and BSW-KU schemes, all elliptic-curve points are represented in compressed format \cite{cohen2005handbook} when they are sent over wireless links.
This allows us to halve their size from 1024 bits to 512 bits.
We further assume a low-bitrate WSAN composed of one gateway, 50 consumers, and 50 producers.
Each consumer is described by an attribute set of 20 attributes.
We assume that the consumer identifiers and the producer identifiers are both 64-bit long.

Table \ref{tab} shows the traffic overhead of consumer leave and producer leave procedures of SEA-BREW and BSW-KU schemes.
\begin{table}
\centering

	\begin{tabular}{c|ccc}
		
		&Size of 		& Number/size of& Total\\
		&broadcast		& unicast		& \\
		&message		    & messages		& \\ 
		&(bytes)         & (bytes)       &  (bytes)\\\hline \hline
		\textbf{SEA-BREW} 		\\
		consumer leave & 252 		    & -				& 252 \\ \hline
		producer leave & 48 		    & -				& 48 \\ \hline
		\textbf{BSW-KU}  & &	        & \\

		consumer leave & 256 		& 50$\times$2,688         & 134,656\\ \hline
		producer leave & 48 		    & -				& 48 \\ \hline
		
	\end{tabular}
	\caption{Traffic overhead of key revocation procedures in the WSAN.}\
	
		\label{tab}

\end{table}
In SEA-BREW, the broadcast message sent by the WSAN gateway during the consumer leave procedure is composed by the ECDSA signature (40 bytes), $U_{EK}$ (64 bytes), and $U_{DK}$ encrypted with the broadcast public key (148 bytes).
Here we assumed that $U_{DK}$ is encrypted by one-time pad with a key encrypted by the Boneh et al.'s broadcast encryption scheme \cite{boneh2005collusion}, so it is composed of 20 bytes (the one-time-padded $U_{DK}$) plus the broadcast encryption overhead (128 bytes).
As can be seen from the table, inside a low-bitrate WSAN, SEA-BREW produces the same traffic overhead as the BSW-KU scheme when performing producer leave procedure.
However, the overhead is merely the 0.2\% of that produced by the BSW-KU scheme when performing a consumer leave procedure.
Indeed, SEA-BREW is able to revoke or renew multiple decryption keys by sending a single 252-byte (considering 80-bit security) broadcast message over the WSAN, opposed to the one 256-byte broadcast message plus 50 unicast messages of 2688-byte each (total: $\sim$131KB of traffic) necessary to update a network with 50 consumers (each of them described by 20 attributes) in a traditional CP-ABE scheme.
With bigger WSANs (more than 50 consumers) or bigger attribute sets (more than 20 attributes) the advantage of SEA-BREW with respect to the BSW-KU scheme grows even more.
Moreover, SEA-BREW also provides a re-encryption mechanism delegated to the untrusted cloud server, which is absent in the BSW-KU scheme.

\subsection{Computational Overhead}

In Table \ref{tab:complex} we compare the computational cost of the primitives of SEA-BREW with those of BSW-KU and of YWRL, in terms of number and type of needed operations.
\begin{table}
\centering
\begin{tabular}{cccc} \hline
Primitive            & Pairings                          & $\mathds{G}_0$ exp.'s                 & $\mathds{G}_1$ exp.'s    \\ \hline
\textbf{SEA-BREW}                                                                                                           \\ \hline
$\mathrm{Encrypt}$      & -                                 & $2|\mathcal{P}|$                      & $1$                   \\
$\mathrm{KeyGen}$  & -                                 & $2|\gamma|+1$                         & -                          \\
$\mathrm{Decrypt}$      &  $2|\mathcal{P}|+1$               & -                                     &$|\mathcal{P}| + 2$    \\
$\mathrm{UpdateCP}$   &  -                                & $1$                                   & -                     \\
$\mathrm{UpdateDK}$      &  -                                & $1$                                   & -                     \\ \hline
\textbf{BSW-KU}                                                                                                     \\\hline 
$\mathrm{Encrypt}$      & -                                 & $2|\mathcal{P}|$                      & $1$                   \\
$\mathrm{KeyGen}$  & -                                 & $2|\gamma|+1$                         & -                     \\
$\mathrm{Decrypt}$      &  $2|\mathcal{P}|+1$               & -                                     &$|\mathcal{P}| + 2$    \\
$\mathrm{UpdateCP}$   & & (not available)                                                                                 \\
$\mathrm{UpdateDK}$      & -                                 & $2|\gamma|+1$                         & -                     \\\hline
\textbf{YWRL} \cite{yu2010achieving}                                                                                 \\\hline
$\mathrm{Encrypt}$      & -                                 & $|\gamma|$                            &$1$                    \\
$\mathrm{KeyGen}$  & -                                 & $|\mathcal{P}|$                       & -                     \\
$\mathrm{Decrypt}$      & $|\mathcal{P}|$                   &   -                                   & $|\mathcal{P}|$       \\
$\mathrm{UpdateCP}$   & -                                 & $|\gamma \cap \mathcal{A}_{rev}|$     & -                     \\
$\mathrm{UpdateDK}$      & -                                 & $|\mathcal{P}\cap \mathcal{A}_{rev}|$ & -                     \\\hline
\end{tabular}
\caption{
Comparison between SEA-BREW, BSW-KU, and YWRL schemes in terms of the computational cost of the primitives. 
For the YWRL scheme, the $\mathrm{UpdateCP}$ and the $\mathrm{UpdateDK}$ primitives correspond respectively to the $\mathrm{AUpdateAtt4File}$ and $\mathrm{AUpdateSK}$ of the original paper.
}
\label{tab:complex}
\end{table}%
In the table, the symbol $\mathcal{A}_{rev}$ indicates the set of attributes that have been revoked, therefore the attributes that need to be updated in ciphertexts and decryption keys.
The symbol $|\mathcal{P}|$ is the number of attributes inside the policy $\mathcal{P}$, and the same applies for $|\gamma|$.
The expression $|\gamma \cap \mathcal{A}_{rev}|$ is the number of attributes belonging to both $\gamma$ and $\mathcal{A}_{rev}$, and the same applies to $|\mathcal{P} \cap \mathcal{A}_{rev}|$.
The operations taken into account are pairings, exponentiations in $\mathds{G}_0$, and exponentiations in $\mathds{G}_1$. 
In all the three schemes, we consider the worst-case scenario for the $\mathrm{Decrypt}$ primitive, which corresponds to a policy with an AND root having all the attributes in $\gamma$ as children. 
This represents the worst case since it forces the consumer to execute the $\mathrm{DecryptNode}$ sub-primitive on every node of the policy, thus maximizing the computational cost.

From the table we can see that SEA-BREW and BSW-KU pay the flexibility of the CP-ABE paradigm in terms of computational cost, especially concerning the $\mathrm{Encrypt}$ and $\mathrm{Decrypt}$ operations.
However, this computational cost is the same of that in Bethencourt et al.'s scheme \cite{bethencourt2007ciphertext}, which has proven to be supportable by mobile devices \cite{ambrosin2015feasibility} and constrained IoT devices \cite{girgenti2019feasibility}.
Note that our $\mathrm{UpdateCP}$ and $\mathrm{UpdateDK}$ primitives have a cost which is independent of the number of attributes in the revoked decryption key.
Such primitives require a single $\mathds{G}_0$ exponentiation, and a number of $\mathds{Z}_p$ multiplications equal to the number of revocations executed from the last update of the ciphertext or the decryption key.
However, the latter operations have a negligible computational cost compared to the former one, therefore we can consider both primitives as constant-time.

Since modern cloud services typically follow a ``pay-as-you-go'' business model, in order to keep the operational costs low it is important to minimize the computation burden on the cloud server itself.
We investigated by simulations the cloud server computation burden of our Lazy PRE scheme compared to the YWRL one, which represents the current state of the art.
We can see from Table \ref{tab:complex} that in both SEA-BREW and YWRL, the cloud performs only exponentiations in $\mathds{G}_0$.

The reference parameters for our simulations are the following ones.
We simulated a system of 100k ciphertexts stored on the cloud server, over an operation period of 1 year.
We fixed an attribute universe of 200 attributes.
We fixed a number of 15 attributes embedded in policies and attribute sets.
We modeled the requests with a Poisson process with average of 50k daily requests.
Finally, we modeled that several consumer leave procedures are executed at different instants, following a Poisson process with average period of 15 days.
In order to obtain more meaningful statistical results we performed 100 independent repetitions of every simulation.

\begin{figure}[ht]
	\centering
	\includegraphics[scale=0.8]{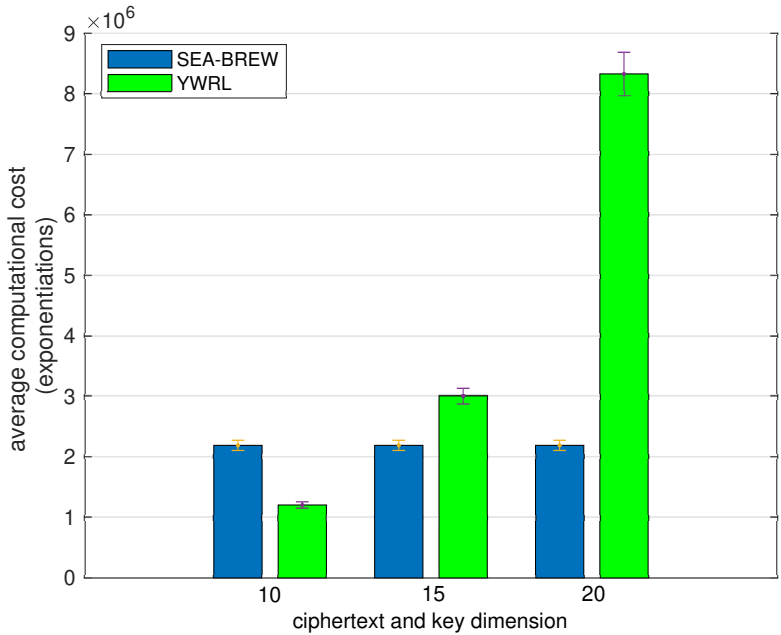}
	\caption{
		Average number of exponentiations over a year, varying policies and attribute sets dimension.
		95\%-confidence intervals are displayed in error bars. 		
	}
	\label{fig:bar1}
\end{figure}
Fig. \ref{fig:bar1} shows the average number of exponentiations in $\mathds{G}_0$ performed by the cloud server, with respect to the number of attributes in ciphertexts and decryption keys, which is a measure of the complexity of the access control mechanism.

As we can see from the figure, SEA-BREW scales better than the YWRL as the access control complexity grows.
This is because in the YWRL scheme every attribute has a singular and independent version number, and the revocation of a decryption key requires to update all the single attributes in the key.
The cloud server re-encrypts a ciphertext with a number of operations equal to the attributes shared between the ciphertext and the revoked key.
Such a number of operations grows linearly with the average number of attributes in ciphertexts and decryption keys.
On the other hand, in SEA-BREW the master key version number is unique for all the attributes, and the revocation of a decryption key requires to update only it.
The cloud server re-encrypts a ciphertext with an operation whose complexity is independent of the number of attributes in the ciphertext and the revoked key.

\begin{figure}[ht]
	\centering
	\includegraphics[scale=0.8]{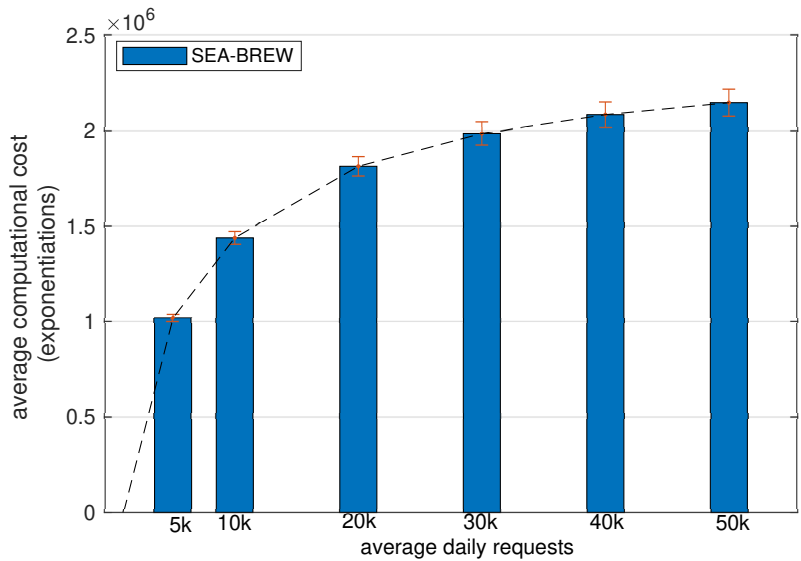}
	\caption{Average number of exponentiation over a year, varying the average daily requests.}
	\label{fig:bar3}
\end{figure}

Fig. \ref{fig:bar3} shows the average number of exponentiations in $\mathds{G}_0$ performed by the cloud server with respect to the average daily requests, which is a measure of the system load.
The number of attributes in ciphertexts and decryption keys is fixed to 15.

Fig. \ref{fig:bar3} shows the average number of exponentiations in $\mathds{G}_0$ performed by the cloud server with respect to the average daily requests, which is a measure of the system load.
The number of attributes in ciphertexts and decryption keys is fixed to 15.
As we can see from the figure the computational load on the cloud server grows sub-linearly with respect to the increase of the requests. 
This behavior allows SEA-BREW to scale well also with high number of requests.
\section{Conclusion}
\label{sec:conclusion}

In this paper, we proposed SEA-BREW (Scalable and Efficient ABE with Broadcast REvocation for Wireless networks), an ABE revocable scheme suitable for low-bitrate Wireless Sensor and Actuator Networks (WSANs) in IoT applications. 
SEA-BREW is highly scalable in the number and size of messages necessary to manage decryption keys.
In a WSAN composed of $n$ decrypting nodes, a traditional approach based on unicast would require $\mathcal{O}(n)$ messages. 
SEA-BREW instead, is able to revoke or renew multiple decryption keys by sending a single broadcast message over a WSAN.
Intuitively, such a message allows all the nodes to locally update their keys.
Also, our scheme allows for per-data access policies, following the CP-ABE paradigm, which is generally considered flexible and easy to use \cite{bethencourt2007ciphertext,liu2013white,ambrosin2015feasibility}.
In SEA-BREW, things and users can exchange encrypted data via the cloud, as well as directly if they belong to the same WSAN.
This makes the scheme suitable for both remote cloud-based communications and local delay-bounded ones.
The scheme also provides a mechanism of proxy re-encryption \cite{yu2010achieving,yu2010attribute,zu2014new} by which old data can be re-encrypted by the cloud to make a revoked key unusable.
We formally proved that our scheme is adaptively IND-CPA secure also in case of an untrusted cloud server that colludes with a set of users, under the generic bilinear group model.
We finally showed by simulations that the computational overhead is constant on the cloud server, with respect to the complexity of the access control policies.

\section*{Funding} 
This work was supported by: 
the European Processor Initiative (EPI) consortium, under grant agreement number 826646; 
the project PRA\_2018\_81 ``Wearable sensor systems: personalized analysis and data security in healthcare'' funded by the University of Pisa;
and the Italian Ministry of Education and Research (MIUR) in the framework of the CrossLab project (Departments of Excellence).\newline

\bibliographystyle{unsrtnat}
\bibliography{biblio.bib}
\end{document}